\documentclass[conference,letterpaper]{IEEEtran}
\usepackage{cite}
\usepackage{amsmath,amssymb,amsfonts}
\usepackage{graphicx}
\usepackage{textcomp}
\usepackage{booktabs}
\usepackage{xcolor}
\usepackage{url}
\usepackage{microtype}
\usepackage{balance}
\usepackage[hidelinks]{hyperref}
\hypersetup{pdftitle={High-Throughput GPU Decoding of Quantum Bivariate Bicycle Codes under Circuit-Level Noise},pdfauthor={Aditya Sodhani and Keshab K. Parhi}}
\graphicspath{{figures/}}
\newcommand{\code}[3]{\ensuremath{[\![#1,#2,#3]\!]}}
\newcommand{\us}{\ensuremath{\,\mu\mathrm{s}}}
\newcommand{\ms}{\ensuremath{\,\mathrm{ms}}}
\newcommand{\MainLER}{\ensuremath{7.58\times10^{-6}}}

\begin{document}
\title{High-Throughput GPU Decoding of Quantum Bivariate Bicycle Codes under Circuit-Level Noise}
\author{\IEEEauthorblockN{Aditya Sodhani and Keshab K. Parhi}
\IEEEauthorblockA{University of Minnesota Twin Cities, Minneapolis, MN 55455 USA\\
Email: \{sodha005, parhi\}@umn.edu}}
\maketitle
\begin{abstract}
Bivariate bicycle (BB) codes reduce the physical-qubit overhead of quantum error correction compared to other codes such as Surface codes, but circuit-level faults create dependencies across measurement rounds that limit decoding throughput. We present a parallel-window GPU decoder that combines reusable fault patterns, compact shared-memory state, and persistent GPU workers using min-sum with Relay-BP recovery. On an RTX~5090, the proposed decoder keeps pace with one syndrome round every 1\us\ across all four evaluated codes, up to \code{144}{12}{12} with 63--144\us\ median commit delays. This rate has not been demonstrated in prior circuit-level GPU decoding studies and it provides $120\times$ higher decoding throughput than sequential sliding-window decoding. Checking neighboring detector rounds before accepting a decoding result reduces failure rates by up to 65\%. For \code{144}{12}{12} at $p=0.002$, our proposed decoder achieves $7.58\times10^{-6}$ failures per round, 36\% lower than the published belief propagation (BP) with syndrome flipping (BP-SF) result.
\end{abstract}
\begin{IEEEkeywords}
Quantum error correction, bivariate bicycle codes, circuit-level noise, GPU architecture, parallel window decoding.
\end{IEEEkeywords}
\section{Introduction}
Physical noise limits the reliability of quantum computing~\cite{preskill2018quantum}, requiring quantum error correction (QEC) to encode logical information across multiple physical qubits~\cite{shor1995scheme,11417923,11695207,sodhani2026noise}. Rather than reading the encoded state directly, QEC repeatedly measures stabilizers~\cite{gottesman1997stabilizer} and streams the resulting syndrome data to a classical decoder to infer and track errors. In superconducting processors, hardware generates a new syndrome round approximately every microsecond~\cite{google2025quantum}. To prevent a backlog that delays computation, a streaming decoder must keep pace with this input rate~\cite{terhal2015quantum}.

Bivariate bicycle (BB) codes~\cite{bravyi2024high} are a promising family of quantum low-density parity-check (QLDPC) codes~\cite{breuckmann2021quantum,panteleev2021quantum} that protect multiple logical qubits using substantially fewer physical qubits than Surface codes~\cite{bravyi2024high,dennis2002topological}. However, circuit-level faults couple syndrome measurements across space and time, making fast and reliable decoding challenging~\cite{gong2024toward}.

GPUs are attractive for QLDPC decoding because they can update many checks and error estimates simultaneously~\cite{wang2011massively,kang2012parallel}. Belief propagation (BP)~\cite{poulin2008iterative} supports these parallel updates, while methods such as BP-OSD~\cite{roffe2020decoding,panteleev2021degenerate}, Relay-BP~\cite{muller2025improved}, and BP-SF~\cite{wang2026fully} help find valid corrections when standard BP fails. A GPU min-sum decoder for data-qubit errors has been presented in~\cite{ferraz2025gpu}; however, their evaluation does not address circuit-level noise or continuous streaming workloads. GPU decoding was extended to process continuous streams by decoding short blocks of consecutive rounds together (sliding windows) in~\cite{wang2026gpu}. However, because their evaluation relies on slow, millisecond-scale intervals, neither study establishes the real-time, $\sim$1\us\ processing required by fast superconducting QEC cycles~\cite{google2025quantum,skoric2023parallel}.

The main limitation of sliding-window decoding is its sequential execution~\cite{gong2024toward,huang2024increasing}. Each window must finish decoding and pass its boundary corrections to the next window before processing can continue. If one window takes longer to decode, it also delays all subsequent windows because they depend on its result. Although GPUs can process multiple independent experiments simultaneously, this does not remove the dependencies within a single stream. To avoid these delays, we must execute multiple windows simultaneously and later reconcile the information passed from one window to the next. Sustaining this approach at microsecond scales requires utilizing the GPU's on-chip shared memory, which is significantly faster than standard global memory, to prevent data-access bottlenecks.

To achieve fast and reliable real-time decoding, our proposed architecture introduces three clear novelties. First, by exploiting the repetitive structure of BB circuits, we utilize a compact set of reusable fault patterns rather than storing the full decoding graph. Combined with compressed message storage~\cite{volkov2020low}, this allows persistent GPU workers~\cite{gupta2012study} to retain their iterative min-sum and Relay-BP recovery~\cite{muller2025improved} state entirely within fast on-chip shared memory. Second, workers decode separated time regions concurrently and resolve the intervening gaps entirely on the GPU without per-window host communication. By eliminating sequential dependencies and host-transfer overheads, this fully device-resident parallel-window~\cite{skoric2023parallel} execution achieves a 120$\times$ throughput improvement over sequential decoding, successfully sustaining the critical 1\us\ syndrome arrival rate for codes up to \code{144}{12}{12}. Third, we introduce a boundary-checking technique that evaluates one extra detector round on each side of the committed core before accepting a result, which reduces logical failure rates by up to 65\%.
\section{BB Codes and Circuit-Level Decoding}
\label{sec:background}
\subsection{BB Codes and Decoder Inputs}
BB codes use repeating check patterns on an $\ell\times m$ grid with wrap-around boundaries, called a torus~\cite{bravyi2024high}. Each code contains $N=2\ell m$ data qubits and $M=\ell m$ stabilizer checks in each check family. We evaluate the \code{72}{12}{6}, \code{90}{8}{10}, \code{108}{8}{10}, and \code{144}{12}{12} codes, where the notation \code{n}{k}{d} denotes $n$ data qubits, $k$ logical qubits, and code distance $d$.

We evaluate a Z-basis quantum memory over $T$ noisy syndrome-extraction rounds and a final data-qubit readout~\cite{bravyi2024high,wang2026fully,wang2026gpu}, accounting for circuit-level faults across gates, idle qubits, resets, and measurements. Because BB code X and Z check families are structurally identical, a Z-basis experiment fully captures the computational workload of one check family. The decoder processes \emph{detector bits}~\cite{gidney2021stim} that flag parity changes between consecutive syndrome measurements, or between the final stabilizers and the data-qubit readout. The $M$ detector bits per round $r$ form a vector $d_r$, which are stacked chronologically to yield the overall observed detector vector $d$.

Stim~\cite{gidney2021stim} converts the noisy circuit into a detector error model (DEM). Each possible fault mechanism has a probability, a detector-flip pattern, and a logical-observable mask. In the DEM matrix $D$, each row represents one individual detector, while each column represents one possible fault mechanism. Thus, the detector bits from one round correspond to $M$ matrix rows. Given observed detector bits $d$, the decoder seeks a likely fault estimate $\hat{e}$ satisfying $D\hat{e} = d$ over $\mathrm{GF}(2)$. Because multiple fault combinations can produce the same detector signature, the decoder must identify a highly likely estimate to correctly apply the associated logical masks and prevent a logical error.

\subsection{Parallel Decoding Windows}
Decoding the entire measurement history at once requires a large decoding graph, increasing both memory usage and computational cost as more rounds accumulate. Sliding-window decoding processes detector bits from $W$ consecutive rounds and finalizes (commits) fault estimates for the oldest $C$ rounds~\cite{gong2024toward,huang2024increasing}. Because a committed fault can flip detectors in the next window, each window must wait for its predecessor's result before decoding.

\begin{figure}[!htb]
\centering
\includegraphics[width=\columnwidth]{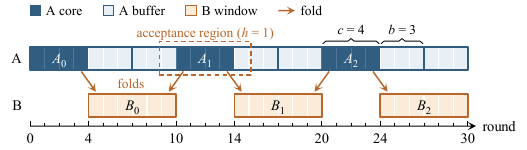}
\caption{Parallel windows ($c=4$, $b=3$): A cores use buffers for context; B gaps wait for neighboring A folds. Dashed: extra acceptance rounds, not extra ownership.}
\label{fig:windows}
\end{figure}

Parallel-window decoding~\cite{skoric2023parallel} breaks this sequential chain by assigning separated regions to A windows and the intervening gaps to B windows (see Fig.~\ref{fig:windows}). Each A window contains a core of $c$ rounds and a buffer of $b$ rounds on each side. Buffers provide context to resolve errors near the core without requiring the A window to own all buffer faults. Each A window commits faults within its core and faults that affect detectors on both sides of a core boundary. Faults entirely within the buffers are left for neighboring B windows. Consecutive core starts are spaced $c+2b$ rounds apart, leaving a $2b$-round gap for each B window.

For example, with $c=4$ and $b=3$, windows $A_0$ (clipped at round 0) and $A_1$ own cores at rounds 0--3 and 10--13, while $B_0$ covers the gap at rounds 4--9. If $A_0$ commits a boundary-crossing fault that flips detectors in rounds 3 and 4, $B_0$ must account for that fixed round-4 flip---called a \emph{fold}---rather than explain it again. A windows can decode independently once their measurements are available. Each B window then uses the boundary corrections from its two neighboring A windows to decode the remaining errors. Since B windows do not depend on one another, multiple B windows can also execute in parallel once their neighboring A windows have finished decoding.
\section{Proposed Decoder}
\label{sec:architecture}
\subsection{Reusing Fault Patterns Across Space and Time}
\label{sec:templates}
Although physical faults occur randomly, the repetitive structure of BB syndrome-extraction circuits across space and time constrains their possible fault patterns to a small, repeating set. We exploit this structure to reuse a small set of patterns instead of storing the full decoding graph across the measurement history. In the evaluated circuits, each DEM column affects detectors in at most two consecutive rounds. We assign each column to its earliest affected round $r$, so its detector flips occur in round $r$ or $r + 1$. All syndrome-extraction rounds except the first and last use the same 10 bulk fault-pattern types. The first extraction, last extraction, and final data readout use 9, 10, and 3 pattern types, respectively.

\begin{figure}[!htb]
\centering
\includegraphics[width=\columnwidth]{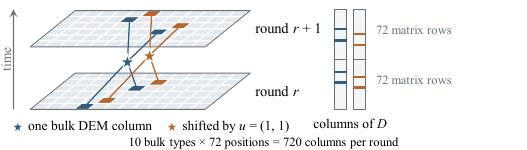}
\caption{A \code{144}{12}{12} bulk DEM mechanism at $p=0.002$ flips two detectors in each of rounds $r$, $r+1$; a shift gives another column.}
\label{fig:torus}
\end{figure}

Each fault-pattern type appears at different positions on the torus (see Fig.~\ref{fig:torus}). Moving a pattern shifts all of its detector positions together, wrapping around the grid edges. For a pattern placed at torus position $u$ and round $r$, a stored spatial offset $v$ and time offset ${\delta \in \{0, 1\}}$ give the affected detector location ${(r + \delta, (u + v) \bmod (\ell, m))}$. Thus, one stored pattern describes connections across many positions and rounds. For example, the \code{144}{12}{12} code has 72 torus positions, so its 10 bulk pattern types generate all 720 DEM columns per round. Error probabilities are bound to the pattern type, but logical-observable masks must be stored per spatial position because they vary under translation. We confirm that expanding these pattern sets identically reproduces the raw DEM connections, probabilities, and masks.

The patterns describe which detectors each fault mechanism affects. During check updates, the decoder needs the reverse lookup: which fault mechanisms can affect each detector. We store this reverse lookup, along with each mechanism's detector list, in compact neighbor tables. Each bulk detector connects to 35 mechanisms: 22 assigned to its own round and 13 to the preceding round. These tables occupy 10.7--21.4\,KiB and are loaded once into each worker's shared memory for reuse across windows. The connectivity is reused across time, while each window maintains its own detector observations and decoding state.

\subsection{Decoding One Window in Shared Memory}
\label{sec:window}
Each worker, a persistent CUDA thread block, decodes one $A$ or $B$ window on a single streaming multiprocessor (SM) using min-sum belief propagation (BP), followed by Relay-BP recovery~\cite{muller2025improved} when needed.

\begin{figure}[!htb]
\centering
\includegraphics[width=\columnwidth]{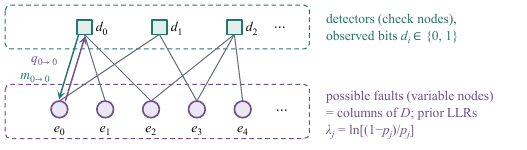}
\caption{Tanner graph of one decoding window.}
\label{fig:tanner}
\end{figure}
\subsubsection{Graph and Priors}
A window is decoded on a Tanner graph (see Fig.~\ref{fig:tanner}) where square check nodes represent the window's detectors and circular variable nodes represent possible circuit faults (DEM columns) rather than data qubits. Edges connect each fault to the detectors it can flip, using the fault patterns and neighbor tables from Section~\ref{sec:templates}. Each fault $j$ starts with a prior log-likelihood ratio (LLR) $\lambda_j=\ln[(1-p_j)/p_j]$ based on its Stim error probability $p_j$.
A positive belief favors no fault ($\hat{e}_j = 0$), whereas a negative belief selects the fault ($\hat{e}_j = 1$).

\subsubsection{Min-Sum Updates}
Each iteration alternates between parallel check updates and parallel variable updates. In the check pass, detector $i$ sends each connected fault $j$ a message $m_{i\rightarrow j}$ computed from its observed bit $d_i \in \{0, 1\}$ and the incoming messages $q_{k\rightarrow i}$ from all \emph{other} connected faults $k \in \mathcal{N}(i)\setminus j$:
\begin{equation}
 m_{i\rightarrow j} = \alpha_t (-1)^{d_i} \prod_{k \in \mathcal{N}(i)\setminus j} \mathrm{sgn}(q_{k\rightarrow i}) \min_{k \in \mathcal{N}(i)\setminus j} \vert{}q_{k\rightarrow i}\vert{},
 \label{eq:check}
\end{equation}
where $\alpha_t = 1 - 2^{-t}$ scales iteration $t$~\cite{volkov2020low,ldpcpkg}. Leaving out fault $j$'s own input prevents sending its own prior evidence back to it. In the variable pass, fault $j$ sums all incoming check messages with its prior $\lambda_j$ to form an updated belief $L_j$, and subtracts detector $i$'s message when sending $q_{j\rightarrow i}$ for the next iteration:
\begin{equation}
 L_j = \lambda_j + \sum_{k \in \mathcal{N}(j)} m_{k\rightarrow j}, \qquad q_{j\rightarrow i} = L_j - m_{i\rightarrow j}.
 \label{eq:var}
\end{equation}

\subsubsection{Compact Shared-Memory State}
Because each check update (\ref{eq:check}) takes the minimum over all incoming magnitudes \emph{except} the recipient's, every outgoing magnitude is either the smallest or second-smallest incoming value. Instead of storing a separate message for every graph edge, each detector stores only a 16-byte summary (the two smallest incoming magnitudes, the minimum's index, the input signs, and the overall parity), and each fault stores only its 32-bit belief $L_j$. Threads reconstruct $m_{i\rightarrow j}$ and $q_{j\rightarrow i}$ on the fly via (\ref{eq:check}) and (\ref{eq:var}) during each pass. Across all four codes, a 10-round $A$ window requires only $32.7\text{--}65.3\text{ KiB}$ of shared memory, including cached neighbor tables, keeping all iterative updates on-chip.

\subsubsection{Fast Path and Recovery}
Each check pass tests whether the current fault estimates reproduce the observed detector bits. For A windows, acceptance requires a match throughout the core and one available buffer round on each side (acceptance margin $h=1$). Checking these extra rounds helps improve boundary decisions without changing which faults are committed. B windows must match all their detector bits after applying the neighboring A-window folds. Most windows finish within a fast path of at most 80 min-sum iterations. If a window is not accepted after 80 iterations, the same worker runs Relay-BP recovery~\cite{muller2025improved}, continuing from the fast-path beliefs for up to 64 legs of 60 iterations with per-column memory weights in $[-0.24, 0.66]$ to escape stuck error patterns. The worker commits its own columns as soon as an accepted estimate is found; if recovery is exhausted, it flags a failure and commits its saved min-sum estimate so dependent windows are not blocked (see Fig.~\ref{fig:architecture}).

\subsection{Keeping Workers Active Without Host Round Trips}
\label{sec:persistent}
\begin{figure*}[t]
\centering
\includegraphics[width=\textwidth]{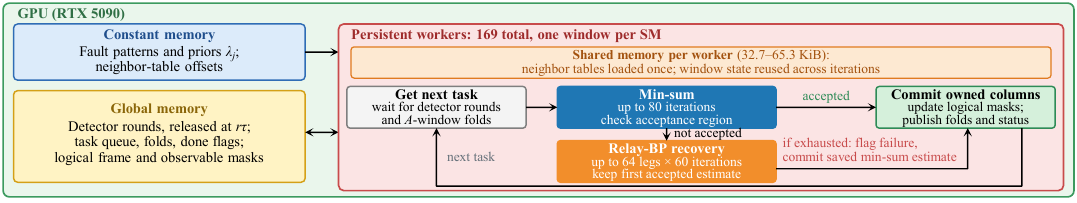}
\caption{Device-resident execution: persistent workers decode, recover, and commit windows on the GPU without host round trips.}
\label{fig:architecture}
\end{figure*}

A single kernel is launched once and keeps its workers active across windows (see Fig.~\ref{fig:architecture}). The default RTX~5090 configuration uses 169 workers and one device-clock block. The host preloads the model, tasks, and detector samples; the clock releases round $r$ at $r\tau$. This emulates arrivals, not live data transfer. An atomic counter on the task queue assigns tasks in the order $A_0$, $A_1$, $B_0$, $A_2$, $B_1$, \ldots\ for each family, but they need not finish in that order. A windows can overlap, while each B window waits for its two A neighbors. Workers may receive tasks before their inputs or dependencies are ready and wait on-device. Before decoding, a B worker applies its neighbors' folds to its observed detector bits $d_B$ and solves $D_B\hat e_B=d_B\oplus f_L\oplus f_R$, where $D_B$ contains only uncommitted fault columns. After decoding, each worker XORs the logical masks of committed faults into the logical frame, publishes folds and completion status, and takes another task without host communication.
\begin{table*}[!t]
\caption{Comparison with Prior Decoders for BB Codes}
\label{tab:compare}
\centering
\scriptsize
\setlength{\tabcolsep}{3pt}\renewcommand{\arraystretch}{0.9}
\begin{tabular}{@{}lllllcclll@{}}
\toprule
 & & & & & \multicolumn{2}{c}{$p_L$/round ($10^{-6}$)} & Compute & Time per & Sustained \\
\cmidrule(lr){6-7}
Decoder & Platform & Noise & Codes ($n$) & Windows & 12 r & 128 r & ($\mu$s/round) & 12-round memory & arrivals \\
\midrule
BP-OSD~\cite{roffe2020decoding} & CPU & CL, $p=0.002$ & 72, 90, 108, 144 & one & 11.4 & -- & 3,200 & 38.4\,ms & -- \\
Min-sum~\cite{ferraz2025gpu} & RTX~4090 & CC & 72--784 (5) & none & -- & -- & 17.1 & -- & -- \\
BP-SF~\cite{wang2026fully,wang2026gpu} & CPU / V100 & CL, $p=0.002$ & 72, 144, 288 & one & 4.38 & -- & -- & 5.5\,ms (V100, $p=0.003$) & -- \\
BP-SF, sliding~\cite{wang2026gpu} & RTX~5060\,Ti & CL, $p=0.002$ & 144, 288 & sliding C7/B3 & -- & 11.9 & 35.4 & -- & 2.97\,ms \\
Proposed, sliding & RTX~5090 & CL, $p=0.002$ & 72, 90, 108, 144 & sliding W10/C7, $h=1$ & -- & \textbf{5.47} & \textbf{21.4} & -- & 120\,$\mu$s \\
Proposed, parallel & RTX~5090 & CL, $p=0.002$ & 72, 90, 108, 144 & parallel c4/b3, $h=1$ & \textbf{4.08} & 7.58 & \textbf{21.4} & 241\,$\mu$s & \textbf{1\,$\mu$s} \\
\bottomrule
\end{tabular}
\par\vspace{2pt}
\parbox{\textwidth}{\scriptsize CL: circuit-level noise; CC: code capacity (data-qubit errors only); --: not reported or not applicable. $p_L$ for \code{144}{12}{12} at $p=0.002$. Compute: mean window service per check-family round.}
\end{table*}
\section{Experimental Results}
\label{sec:evaluation}
\subsection{Experimental Setup}
We evaluate the four BB codes on an NVIDIA RTX~5090 (CUDA~12.6) using the standard depth-eight circuit-level noise model~\cite{bravyi2024high} at physical error rate $p=0.002$, which injects faults across CNOTs, idle layers (1, 7, and 8), resets, and measurements. Unless noted, parallel windows use $c=4$, $b=3$, and acceptance margin $h=1$.

We evaluate decoding accuracy and streaming throughput separately. Accuracy benchmarks run $2\times10^{4}$ to $3\times10^{7}$ independent $T$-round Z-basis memory experiments per data point. An experiment fails if any window remains unaccepted after recovery or if the final logical readout is incorrect, yielding a normalized per-round logical failure rate $p_L=1-(1-N_f/N_s)^{1/T}$. To benchmark full-code streaming throughput without joint X/Z correlation tracking, we run two concurrent Z-basis syndrome streams (three seeds of $2\times10^{6}$ preloaded rounds per stream) to model the combined workload of both check families. A streaming run passes if the mean commit delay---measured from scheduled round arrival to window commit---grows by no more than 25\% plus $5\us$ between the first and final 10\% of rounds.

\subsection{Decoding Accuracy}
Boundary checking significantly reduces windowing errors. Across 128-round parallel-window experiments, enforcing a one-round acceptance margin ($h=1$) reduces per-round failure rates by $1.9$--$2.9\times$ (up to 65\%) on the three larger codes compared to core-only acceptance ($h=0$), reaching $p_L=\MainLER$ on \code{144}{12}{12}.

Min-sum with Relay-BP outperforms BP-OSD on larger codes. For $d$-round single-window experiments, our proposed decoder achieves $1.6$--$2.8\times$ (39--64\%) lower failure rates than CPU BP-OSD (\texttt{ldpc}~2.4.1, 1,000 adaptive min-sum iterations, OSD-CS order~10) on the three larger codes, while showing a $1.4\times$ higher rate on \code{72}{12}{6}. Sweeping physical noise from $p=2\times10^{-3}$ down to $2\times10^{-4}$ over 128 rounds confirms that all four codes remain well below the $p_L=p$ pseudothreshold, with larger codes suppressing logical errors steeply down to $p_L\approx2.6\times10^{-10}$ per round for \code{144}{12}{12} at $p=2\times10^{-4}$.
\subsection{Streaming Throughput and Multi-Code Capacity}
Parallel windows unlock $1\us$ real-time streaming---a $120\times$ throughput gain over sequential windows. All four evaluated codes pass the $2\times10^{6}$-round streaming stability test at the target $\tau=1\us$ arrival interval (with mean-delay drift under 6.1\%) and sustain arrivals $1.7$--$3.3\times$ faster in stress sweeps. On \code{144}{12}{12}, our sequential sliding-window baseline ($W=10$, $C=7$, $h=1$) stalls below $\tau=120\us$, whereas parallel windows sustain $\tau=1\us$ ($0.6\us$ shortest passing interval) despite similar per-round computation. Overlapping independent windows allows subsequent regions to progress on other SMs while a difficult window undergoes Relay-BP recovery.

A single GPU supports up to 72 logical qubits at $1\us$ rounds. In 200,000-round capacity sweeps at $\tau=1\us$ using two 512-thread workers per SM for the two smallest codes, one RTX~5090 concurrently sustains 72, 32, 16, and 12 modeled logical qubits for the \code{72}{12}{6}, \code{90}{8}{10}, \code{108}{8}{10}, and \code{144}{12}{12} codes, respectively. Across codes, median commit delays remain tight at $63$--$144\us$.

\subsection{Comparison with Prior Decoders}
Table~\ref{tab:compare} compares our architecture against state-of-the-art BB decoders on \code{144}{12}{12}. In 12-round single-window decoding, our GPU decoder reduces logical failure rates by 64\% over BP-OSD~\cite{roffe2020decoding} and 7\% over BP-SF~\cite{wang2026fully,wang2026gpu}, while cutting average execution latency to $241\us$---a $22.8\times$ speedup over BP-SF on a V100 ($5.5\ms$). In 128-round streaming, our sequential sliding-window decoder achieves a 54\% lower failure rate than BP-SF ($5.47\times10^{-6}$ vs.\ $11.9\times10^{-6}$), while our parallel-window decoder simultaneously improves accuracy by 36\% ($7.58\times10^{-6}$) and increases sustained arrival throughput by more than three orders of magnitude ($1\us$ vs.\ $2.97\ms$).
\section{Conclusion}
\label{sec:conclusion}
We presented a parallel-window GPU decoder for BB codes under circuit-level noise that combines reusable fault patterns, compact shared-memory state, and persistent workers. Boundary checking reduces logical failure rates by up to 65\%, while parallel windows sustain $1\us$ syndrome arrivals through \code{144}{12}{12}---a $120\times$ throughput gain over sequential decoding. Future work will extend this architecture to correlated X/Z decoding and fault-tolerant logical operations.
\clearpage
\balance
\bibliographystyle{IEEEtran}
\bibliography{refs}
\end{document}